\UseRawInputEncoding

\documentclass[prl,twocolumn,showpacs,groupedaddress,superscriptaddress,nofootinbib,floatfix,preprintnumbers]{revtex4-1}
\usepackage{amssymb,amsmath,graphicx,color}
\usepackage{bm}
\usepackage[tight]{subfigure}
\usepackage[export]{adjustbox}
\usepackage{braket}
\usepackage[colorlinks=true,citecolor=blue,linkcolor=blue]{hyperref}
\usepackage[table]{xcolor}
\usepackage{cancel}
\usepackage{multirow}

\newenvironment{psmallmatrix}
  {\left(\begin{smallmatrix}}
  {\end{smallmatrix}\right)}

\begin{document}

\title{
The path from lattice QCD to the short-distance contribution
\\
to $0\nu\beta\beta$ decay with a light Majorana neutrino
}

\author{Zohreh Davoudi{\footnote{\tt davoudi@umd.edu}}
}
\affiliation{Maryland Center for Fundamental Physics and Department of Physics, 
University of Maryland, College Park, MD 20742, USA}
\affiliation{
RIKEN Center for Accelerator-based Sciences,
Wako 351-0198, Japan}

\author{Saurabh V. Kadam{\footnote{\tt ksaurabh@umd.edu }}
}
\affiliation{Maryland Center for Fundamental Physics and Department of Physics, 
University of Maryland, College Park, MD 20742, USA}

\preprint{UMD-PP-020-10}

\begin{abstract}
Neutrinoless double-$\beta$ ($0\nu\beta\beta$) decay of certain atomic isotopes, if observed, will have significant implications for physics of neutrinos and models of physics beyond the Standard Model. In the simplest scenario, if the mass of the light neutrino of the Standard Model has a Majorana component, it can mediate the decay. Systematic theoretical studies of the decay rate in this scenario, through effective field theories matched to \emph{ab initio} nuclear many-body calculations, are needed to draw conclusions about the hierarchy of neutrino masses, and to plan the design of future experiments. However, a recently identified short-distance contribution at leading order in the effective field theory amplitude of the subprocess $nn \to pp\,(ee)$ remains unknown, and only lattice quantum chromodynamics (QCD) can directly and reliably determine the associated low-energy constant. While the numerical computations of the correlation function for this process are underway with lattice QCD, the connection to the physical amplitude, and hence this short-distance contribution, is missing. A complete framework that enables this complex matching is developed in this Letter. The complications arising from Euclidean and finite-volume nature of the corresponding correlation function are fully resolved, and the value of the formalism is demonstrated through a simple example. The result of this work, therefore, fills the gap between first-principles studies of the $nn \to pp\,(ee)$ amplitude from lattice QCD and those from effective field theory, and can be readily employed in the ongoing lattice-QCD studies of this process.
\end{abstract}

\maketitle

{\it Introduction.---}The lepton-number violating process $(A,Z) \to (A,Z+2)+ee$, with $A$ and $Z$ being, respectively, the atomic and proton numbers of a parent nucleus, if observed, will mark a major discovery. Beyond its confirmation of the presence of a Majorana component to the neutrino mass~\cite{Schechter:1981bd}, our knowledge of beyond-SM (BSM) mechanisms that may be responsible for this decay can be enhanced by combining theoretical calculations of the rate, and other decay observables, with experimental findings~\cite{Dolinski:2019nrj, DellOro:2016tmg, Bilenky:2014uka}. Furthermore, planned experimental endeavors will crucially benefit from theoretical predictions of the expected rates in various isotopes given the BSM scenarios considered~\cite{Dolinski:2019nrj, Cappuzzello:2018wek, DellOro:2016tmg, Faessler:2012ku, giuliani2012neutrinoless}. A widely considered scenario is a minimal extension of the SM in which the light neutrinos of the SM are promoted to Majorana neutrinos, which by virtue of being their own anti-particles, can be emitted and reabsorbed by the nucleus undergoing the decay. The corresponding nuclear matrix element is long range in nature and receives contributions from intermediate nuclear states. Despite the long-range nature of the process, recent nuclear effective field theory (EFT) analyses of the elementary subprocess $nn \to pp\,(ee)$ have revealed a short-distance contribution to the amplitude at leading order (LO), with a low-energy constant (LEC) of the corresponding isotensor contact operator that absorbs the ultraviolet (UV) scale dependence of the amplitude through Renormalization Group (RG)~\cite{Cirigliano:2017tvr, Cirigliano:2018hja, Cirigliano:2019vdj}. As such a subprocess cannot be observed in free space, and given the program that has been formed around the use of nuclear EFTs to systematically improve the \emph{ab initio} nuclear structure calculations of the nuclear matrix elements~\cite{Menendez:2011qq, Pastore:2017ofx, Basili:2019gvn, Yao:2019rck, Yao:2020olm} toward experimentally-relevant isotopes, the unknown value of such a short-distance contribution appears to impede progress, and has promoted several estimations based on the connection to charge-invariance breaking contribution to two-nucleon scattering~\cite{Cirigliano:2019vdj}, the use of Cottingham formula in the NN sector~\cite{Cirigliano:2020dmx}, and large-$N_c$ considerations~\cite{Richardson:2021xiu}, with varying uncertainties.

Lattice QCD (LQCD), which numerically solves QCD on a finite grid in an Euclidean spacetime, has the promise of reliably constraining the EFTs of $0\nu\beta\beta$ in the few-nucleon sector~\cite{Cirigliano:2020yhp, Davoudi:2020ngi, Cirigliano:2019jig}, and has already demonstrated its reach and capability in constraining pionic matrix elements for lepton-number violating processes $\pi^- \to \pi^+\,(ee)$ and $\pi^-\pi^- \to ee$ within the light-neutrino scenario~\cite{Feng:2018pdq, Tuo:2019bue, Detmold:2020jqv}, the $\pi^- \to \pi^+\,(ee)$ process within a heavy-scale scenario~\cite{Nicholson:2018mwc}, as well as the (lepton-number conserving) two-neutrino double-$\beta$ decay ($2\nu\beta\beta$) of a two-nucleon state~\cite{Shanahan:2017bgi,Tiburzi:2017iux} (the latter yet at unphysically large quark masses due to the computational cost). LQCD matrix elements for these processes, however, lack certain complexities compared with the desired $nn \to pp\,(ee)$ process with a light Majorana neutrino, whose determination is the key to matching to an EFT description. While the numerical evaluations of the matrix elements are underway, the interpretation of these matrix elements in terms of the physical amplitude, and their matching to EFTs have so far been missing from the course of developments. In this Letter, such a framework will be developed and presented for the first time. This framework, along with a realistic example to be outlined, demonstrate how the results of this work can be used in the upcoming studies to obtain the short-distance LEC of the EFT from LQCD. This matching framework builds upon major developments in recent years in accessing local and non-local transition amplitudes in hadronic physics from the corresponding finite-volume matrix elements in Euclidean spacetime obtained with LQCD~\cite{Lellouch:2000pv, Detmold:2004qn, Meyer:2011um, Briceno:2012yi, Bernard:2012bi, Briceno:2014uqa, Briceno:2015csa, Briceno:2015tza, Christ:2015pwa, Briceno:2019opb, Feng:2020nqj}, and in particular, a recent work on developing a similar formalism for the two-neutrino process $nn \to pp\,(ee\bar{\nu}_e\bar{\nu}_e)$~\cite{Davoudi:2020xdv}. Nonetheless, the neutrinoless process involves additional complexities due to a propagating neutrino in the intermediate state, requiring new components to be included in the matching condition between finite and infinite-volume matrix elements, as well as Minkowski and Euclidean matrix elements. 

\vspace*{3mm}
{\it EFT amplitude at leading order.---}In a SM EFT of $0\nu\beta\beta$ decay~\cite{Weinberg:1979sa, Babu:2001ex, Prezeau:2003xn, deGouvea:2007qla, Lehman:2014jma, Graesser:2016bpz, Cirigliano:2017djv, Cirigliano:2018yza}, the lepton-number (L) violating operator with the lowest mass dimension is a Majorana mass term, $\mathcal{L}^{(\Delta L = 2)}_\nu=-\frac{m_{\beta\beta}}{2} \nu_L^TC\nu_L+{\rm h.c.}$ Here, $C=i\gamma_2\gamma_0$ denotes the charge conjugation matrix, $\nu_L$ is the left-handed (electron) neutrino field, $m_{\beta\beta}=\sum_i U_{ei}^2m_i$ is the effective neutrino mass, with $U_{ei}$ being the elements of the Pontecorvo-Maki-Nakagawa-Sato (PMNS) matrix~\cite{Pontecorvo:1957qd, Maki:1962mu}. $m_i$ is the mass of the neutrino mass eigenstate $i$. While the $0\nu\beta\beta$ decay can only proceed in certain nuclear media, the subprocess to be studied is $nn \to pp\,(ee)$. Since quarks are bound to nucleons and nucleons interact via the non-perturbative strong force, to relate the rate of the decay to the underlying SM EFT, one needs to map this problem to a nuclear EFT, and constrain the EFT, e.g., using a direct calculation of the matrix element with LQCD.

The nuclear EFT  considered here is the pionless EFT~\cite{Kaplan:1998tg, Kaplan:1998we, vanKolck:1998bw, Chen:1999tn, Butler:1999sv}, where the Lagrangian of free and strongly interacting nucleons can be organized as
\begin{equation}
\mathcal{L}_N^{\rm (QCD)}=N^{\dagger }\bigg(i\partial_{t}+\frac{{\nabla}^2}{2M}\bigg)N
-C_{0}(N^{T}P_{i}N)^{\dagger }(N^{T}P_{i}N)+\cdots
\label{eq:Lstrong}
\end{equation}
Here, $\partial_t$ is the time derivative and ${\bf \nabla}$ is the spatial gradient operator. $N=\begin{psmallmatrix} p \\ n \end{psmallmatrix}$ is an isospin doublet comprised of the proton, $p$, and the neutron, $n$, fields, each with mass $M$. Isospin symmetry will be assumed throughout. $P_{i} \equiv \frac{1}{\sqrt{8}}\sigma _{2}\tau _{2}\tau_{i}$ is a projector for the isotriplet channel, and the ellipsis denotes higher-order terms in a momentum expansion. A similar interacting term can be written for the isosinglet channel. The effective Lagrangian for the charged-current (CC) weak interaction is given by
\begin{equation}
{\cal L}_N^{(\rm CC)} =-\frac{4V_{ud}G_{F}}{2\sqrt{2}} \big[\overline{e}_L\gamma ^{\mu }\nu_L\big]  \big[ N^\dagger \tau _{+}(v_\mu-2g_{A} \mathcal{S}_\mu)N\big] +{\rm h.c.} ,
\label{eq:LCC}
\end{equation}
where $G_F$ is Fermi's constant, $V_{ud}$ is a Cabibbo-Kobayashi-Maskawa (CKM) matrix element~\cite{Cabibbo:1963yz, Kobayashi:1973fv}, $v$ and $\mathcal{S}$ are the nucleon velocity and spin, respectively ($v=(1,\bm{0})$ and $\mathcal{S}=(0,\tfrac{\bm{\sigma}}{2})$ in the nucleon's rest frame), $\tau_{+}=(\tau_{1}+i\tau_{2})/2$ where $\tau_i$ are isospin Pauli metrices, and $g_A$ is the nucleon's axial charge. The leptonic current contains the left-handed electron, $e_L$, and neutrino, $\nu_L$, fields. Last but not least, one can construct a contact $\Delta L = 2$ four-nucleon-two-electron operator in the EFT:
\begin{eqnarray}
&&{\cal L}^{(\Delta L = 2)}_N =\bigg(\frac{4V_{ud}G_{F}}{\sqrt{2}}\bigg)^2
m_{\beta\beta} \, g_\nu^{NN} \times
\nonumber\\
&& \hspace{1.5 cm}  \left[\bar{e}_LC\bar{e}_L^T\right]
 \big[(N^TP_-N)^{\dagger }(N^{T}P_+N)\big]+{\rm h.c.},
\label{eq:LNdeltaL2}
\end{eqnarray}
where $P_{+} = (P_{1}+iP_{2})/{2}$~\cite{Cirigliano:2017tvr, Cirigliano:2018hja, Cirigliano:2019vdj}. While naive dimensional analysis suggests that this operator must contribute at a high order, RG considerations require promoting this operator to LO~\cite{Cirigliano:2018hja, Cirigliano:2019vdj}, as will be discussed shortly.

\begin{figure}[t!]
\centering
\includegraphics[width=\columnwidth]{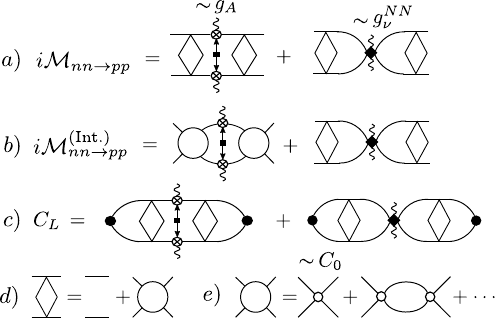}
\caption{Diagrams representing a) the full LO transition amplitude, b) the amplitude excluding neutrino exchanges on the external legs as expressed in Eq.~(\ref{eq:MICnnpp}), and c) the finite-volume correlation function defined in Eq.~(\ref{eq:CL}). The black filled circles correspond to interpolating operators for the initial and final isotriplet states, the solid lines are nucleon propagators, the line with a solid black square denotes the Majorana neutrino propagator, and the wavy lines represent the endpoint of the currents. The ellipsis in e) denotes the chain of s-channel two-nucleon loops connected via $C_0$ couplings.
} 
\label{fig:diagrams}
\end{figure}

The full transition amplitude for the $nn \to pp\,(ee)$ process is not separable to the hadronic and leptonic amplitudes given the presence of a neutrino that propagates between the two weak currents. Nonetheless, the contribution from final-state electrons (as well as constants proportional to $G_F$ and $V_{ud}$) can still be separated from a hadronic amplitude that includes the hadronic matrix element convoluted by the neutrino propagator. This latter contribution is what one would evaluate in LQCD and match to nuclear EFTs. We assume a simple kinematic in which the total three-momenta of the system is zero, and the electrons are at rest, each having energy $E_1=E_2=m_e$, where $m_e$ is the electron's mass. Furthermore, at LO in the EFT, two further simplifications arise: i) only s-wave interactions of the nucleons contribute, ii) the amplitude receives contributions from a static neutrino potential only, and contributions from the small non-zero neutrino mass in the denominator of the neutrino propagator, as well as radiative neutrinos, can be ignored. The mixed hadronic-leptonic amplitude can then be written as $\mathcal{M}_{nn\to pp} \equiv \mathcal{M}^{(\rm{Ext.})}_{nn\to pp}+\mathcal{M}^{(\rm{Int.})}_{nn\to pp}$. $\mathcal{M}^{(\rm{Ext.})}_{nn\to pp}$ denotes contributions in which the neutrino propagates between two external nucleons, see Fig.~\ref{fig:diagrams}. On the other hand, $\mathcal{M}^{(\rm{Int.})}_{nn\to pp}$ denotes contributions in which the neutrino propagates between two nucleons dressed by strong interactions on both sides, as shown in Fig.~\ref{fig:diagrams}.  It is this amplitude that depends upon the short-distance LEC $g_{\nu}^{NN}$ through:
\begin{eqnarray}
&&\mathcal{M}^{(\rm{Int.})}_{nn\to pp} (E_i,E_f)=m_{\beta\beta}\;\mathcal{M}(E_f) \bigg [ -(1+3g_A^2)\times 
\nonumber
\\
%
%
&&\hspace{2.2 cm } J^{\infty}(E_i,E_f;\mu)+\frac{2g_\nu^{NN}(\mu)}{C_0^2(\mu)}\bigg]
\mathcal{M}(E_i),
\label{eq:MICnnpp}
\end{eqnarray}
and will be the subject of matching to LQCD. Here, $E_i$ and $E_f$ denote the energy of the incoming two-neutron state and the outgoing two-proton state, respectively. $\mathcal{M}$ is the LO strong-interaction scattering amplitude of the isotriplet channel,
\begin{eqnarray}
\mathcal{M}(E)=\frac{-1}{C_0^{-1}(\mu)+\tfrac{M}{4\pi}(\mu+\sqrt{ME})},
\label{eq:M}
\end{eqnarray}
and $J^\infty$ is a function representing the s-channel two-loop diagram with an exchanged Majorana neutrino. The two-loop integral is divergent in the UV and in the dimensional regularization scheme is regularized to
\begin{equation}
J^{\infty}(E_i,E_f;\mu)=\frac{M^2}{32\pi^2} \bigg[-\gamma_E+\ln(4\pi)+L(E_i,E_f;\mu) \bigg],
\label{eq:Jinfty}
\end{equation}
where $L(E_i,E_f;\mu) \equiv \ln\left(\tfrac{\mu^2/M}{-(\sqrt{E_i}+\sqrt{E_f})^2-i\epsilon}\right)+1$~\cite{Cirigliano:2018hja, Cirigliano:2019vdj}. The UV divergence of the loop function necessitates introduction of a counterterm at the same order, i.e., $g_\nu^{NN}$. $\mu$ in these equations is a UV renormalization scale, and the requirement of the independence of physical amplitudes $\mathcal{M}$ and $\mathcal{M}^{(\rm{Int.})}_{nn\to pp}$ on such a scale provides RG-flow equations for the LECs $C_0$ and $g_{\nu}^{NN}$, respectively. It should be noted that it is the scale-independent combination $-(1+3g_A^2)J^{\infty}+2g_\nu^{NN}C_0^{-2}$ that can be constrained with LQCD. $g_\nu^{NN}(\mu)$ can then be determined using the values of $C_0$ and $J^{\infty}$ at a given $\mu$.

\vspace*{3mm}
{\it Matching between finite and infinite volume.---}Keeping the Minkowski signature of spacetime intact, we now consider a finite spatial volume with cubic geometry and with extent $L$ along each Cartesian coordinate with periodic boundary conditions. The time direction is assumed to be infinite. Since the amplitudes cannot be defined in a finite-volume, one needs to resort to defining a correlation function instead. At LO in the EFT (see Fig.~\ref{fig:diagrams}):
\begin{eqnarray}
&&C_L(E_i,E_f) = C_{\infty}(E_i,E_f) +
\mathcal{B}_{pp}(E_f)\,i\mathcal{F}(E_f)
\nonumber
\end{eqnarray}
\begin{eqnarray}
&&~~~~\bigg[i\mathcal{M}^{(\rm{Int.})}_{nn\to pp} (E_i,E_f) + m_{\beta\beta}(1+3g_A^2) \, i\mathcal{M}(E_f)
\nonumber\\
&&~~~~~~i\,\delta J^V(E_f,E_i)\, i\mathcal{M}(E_i) \bigg]
i\mathcal{F}(E_i) \mathcal{B}^\dagger_{nn}(E_i)+\cdots.
\label{eq:CL}
\end{eqnarray}
$\mathcal{B}^\dagger_{nn}$ and $\mathcal{B}_{pp}$ are the matrix elements of initial- and final-state interpolating operators between vacuum and on-shell ``in'' and ``out'' two-nucleon states, respectively. The ellipsis denotes terms that will not matter for the matching relation, see Ref.~\cite{Davoudi:2020xdv} for further detail on a similar process. $\mathcal{F}$ is a finite-volume function defined as
\begin{equation}
\mathcal{F}^{-1}(E)=F^{-1}(E)+\mathcal{M}(E),
\label{eq:calF0}
\end{equation}
with $F(E) = \bigg[\frac{1}{L^3}\sum_{\bm{k}  \in \frac{2\pi}{L}\mathbb{Z}^3}-\int\frac{d^3\bm{k}}{(2\pi)^3}\bigg]\frac{1}{E-\frac{{\bm k}^2}{M}+i\epsilon}$. The discretized energy eigenvalues of the two-nucleon system in a finite volume, $E_m$, are obtained from the ``quantization condition'' $\mathcal{F}^{-1}(E_m)=0$~\cite{Luscher:1986pf,Luscher:1990ux}. Finally, a new finite-volume function $\delta J^V$, corresponding to the two-loop diagram with the exchanged neutrino propagator needs to be evaluated:
\begin{eqnarray}
&&\delta J^V(E_i,E_1,E_f)=
\bigg[\frac{1}{{L^6}}\sum_{\substack{\bm{k}_1,\bm{k}_2 \\ \bm{k}_1 \neq \bm{k}_2}}-
\int \frac{d^3k_1}{(2\pi)^3}\frac{d^3k_2}{(2\pi)^3}
\bigg]
\nonumber\\
&&\hspace{1.6 cm}
\frac{1}{E_i-\tfrac{\bm{k}_1^2}{M}+i\epsilon} \frac{1}{E_f-\tfrac{\bm{k}_2^2}{M}+i\epsilon}\frac{1}{|\bm{k}_1-\bm{k}_2|^2},
\label{eq:deltaJV}
\end{eqnarray}
where in the summations, $\bm{k}_1,\bm{k}_2  \in \frac{2\pi}{L}\mathbb{Z}^3$. This sum-integral difference can be evaluated numerically for given values of $E_i$ and $E_f$, the detail of which is presented in Supplemental Material. The requirement $\bm{k}_1 \neq \bm{k}_2$ removes the zero spatial-momentum mode of the neutrino in the loop to render the finite-volume sum finite. Correspondingly, the finite-volume correlation function in LQCD will need to implement a zero-mode regulated neutrino propagator to match to this expression. Such a treatment of the infrared singularities in a finite volume is customary in the lattice QCD+QED studies of hadronic masses~\cite{Hayakawa:2008an, Borsanyi:2014jba, Davoudi:2018qpl}, decay amplitudes~\cite{Lubicz:2016xro, Carrasco:2015xwa, Cai:2018why}, and two-hadron scattering~\cite{Beane:2014qha, Beane:2020ycc}.

To proceed with finding the matching relation, one notes that the finite-volume correlation function in Eq.~(\ref{eq:CL}) has the same general structure as that for the two-neutrino process obtained in Ref.~\cite{Davoudi:2020xdv}. As a result, all steps introduced in Ref.~\cite{Davoudi:2020xdv} can be closely followed to obtain the matching relation between finite and infinite-volume matrix elements. In particular, upon Fourier transforming Eq.~(\ref{eq:CL}) with $E_i$ and $E_f$ to form the correlation function in the mixed time-momentum representation, and comparing it against the same correlation function that is obtained from a direct four-point function upon inserting complete sets of intermediate finite-volume states between the currents, one arrives at
\begin{align}
&L^6\;\bigg|\mathcal{T}^{(\rm M)}_L \bigg|^2=\bigg|\mathcal{R}(E_{n_f})\bigg | \, \bigg|\mathcal{M}^{(\rm{Int.})}_{nn\to pp} (E_{n_i},E_{n_f})-m_{\beta\beta}
\nonumber\\
&\hspace{0.0 cm}(1+3g_A^2)\mathcal{M}(E_{n_f})\delta J^V(E_{n_f},E_{n_i}) \mathcal{M}(E_{n_i}) \bigg|^2 \bigg|\mathcal{R}(E_{n_i})\bigg|,
\label{eq:IVFVmatching}
\end{align}
where $\mathcal{R}(E_n) \,=\, \lim_{E \to E_n} (E-E_n)\;\mathcal{F}(E)$, and $E_{n_{i(f)}}$ denotes a finite-volume energy of the initial (final) two-nucleon state. $\mathcal{T}^{(\rm M)}_L$ denotes the Minkowski finite-volume matrix element defined as
\begin{eqnarray}
&&\mathcal{T}^{(\rm M)}_L \equiv \int dz_0\, e^{iE_1 z_0}\int_L d^3z
\nonumber\\
&&~~\big[\langle E_{n_f},L|\, T[\mathcal{J}(z_0,\bm{z})\,S_\nu(z_0,\bm{z})
\mathcal{J}(0)]\, |E_{n_i},L\rangle\big]_L.
\label{eq:TML}
\end{eqnarray}
Here, $\mathcal{J}=\bar{q} \tau _+\gamma_\mu(1-\gamma_5)q$ with $q=\begin{psmallmatrix} u \\ d \end{psmallmatrix}$, which can be implemented in LQCD calculations. At the hadronic level, it matches to $N^\dagger \tau _{+}(v_\mu-2g_{A} \mathcal{S}_\mu)N$ in Eq.~(\ref{eq:LCC}). Nonetheless, being a quark-level current means that $\mathcal{T}^{(\rm M)}_L$ also incorporates the contact $\Delta L=2$ interaction in Eq.~(\ref{eq:LNdeltaL2}). $S_\nu(z_0,\bm{z})$ denotes the Minkowski finite-volume propagator of a Majorana neutrino, with its zero spatial-momentum mode removed.

\vspace*{3mm}
{\it Minkowski to Euclidean matching.---}The quantity $\mathcal{T}^{(\rm M)}_L$ in Eq.~(\ref{eq:TML}), whose connection to the physical amplitude was established in Eq.~(\ref{eq:IVFVmatching}), is defined with a Minkowski signature. On the other hand, with LQCD only Euclidean correlation functions can be evaluated. Unfortunately in the case of non-local matrix elements, generally one cannot obtain the former from the latter upon an analytical continuation~\cite{Briceno:2019opb}. To appreciate the subtlety involved, and to introduce a procedure that, nonetheless, allows constructing the Minkowski matrix element from its counterpart in Euclidean spacetime, one should consider a correlation function:
\begin{eqnarray}
G_L^{(\rm E)}(\tau)&=&\int_L d^3z \; \big [\langle E_f,L|T^{(\rm E)} [\mathcal{J}^{(\rm E)}(\tau,\bm{z})S_\nu^{(\rm E)}(\tau,\bm{z})
\nonumber\\
&& \hspace{3 cm} \mathcal{J}^{(\rm E)}(0)] |E_i,L \rangle \big]_L,
\label{eq:GLE}
\end{eqnarray}
that can be computed directly with LQCD. $\tau \equiv iz_0$ is the Euclidean time, and the superscript (E) is introduced on Euclidean quantities. In particular, $S_\nu^{(\rm E)}$ is the Euclidean neutrino propagator in a finite volume with its zero spatial-momentum mode removed,
\begin{eqnarray}
&&S_\nu^{(\rm E)}(\tau,\bm{z}) 
= \frac{1}{L^3}\sum_{\bm{q}  \in \frac{2\pi}{L}\mathbb{Z}^3\neq \bm{0}} \int \frac{dq_0^{\rm (E)}}{2\pi}e^{i\bm{q} \cdot \bm{z}-iq_0^{\rm (E)}\tau} \frac{m_{\beta\beta}}{q_0^{\rm (E)2}+|\bm{q}|^2}
\nonumber\\
&& \hspace{0.5 cm} 
=\frac{m_{\beta\beta}}{2L^3}\sum_{\bm{q}  \in \frac{2\pi}{L}\mathbb{Z}^3\neq \bm{0}}\frac{e^{i\bm{q} \cdot \bm{z}}}{|\bm{q}|}\bigg[\theta(\tau)e^{-|\bm{q}| \tau}+\theta(-\tau)e^{|\bm{q}| \tau} \bigg].
\end{eqnarray}
It is now clear that simply integrating over the Euclidean time with weight $e^{E_1 \tau}$ can be problematic if on-shell intermediate states are allowed. Here, $E_1$ is the energy of the first or the second electron depending on the time ordering. This can be seen by expressing the Heisenberg-picture operator in Euclidean spacetime as $\mathcal{J}^{(\rm E)}(\tau,\bm{z})=e^{\hat{P}_0\tau -i \hat{\bm{P}}\cdot \bm{z}} \, \mathcal{J}^{(\rm E)}(0) \,$ $ e^{-\hat{P}_0\tau+i \hat{\bm{P}}\cdot \bm{z}}$, where $\hat{P}_0$ and $\hat{\bm{P}}$ are energy (Hamiltonian) and momentum operators, respectively, and upon inserting a complete set of single- and multi-particle states between the two currents. Without loss of generality, we assume that $\mathcal{J}^{(\rm E)}(0)$ is the same as its Minkowski counterpart. The Euclidean superscript of the Schr\"odinger-picture currents will therefore be dropped. It then becomes clear that for those values of intermediate-states energies and momenta such that  $i) \; |\bm{P}_{*m}|+E_{*m} \leq E_f+E_1~\text{or}~ii)\; |\bm{P}_{*m}|+E_{*m} \leq E_i-E_1$, the integration over Euclidean time with $e^{E_1 \tau}$ will be divergent. Here, $E_{*m}$ are the finite-volume energy eigenvalues of the intermediate spin-triplet two-nucleon state with total momentum ${\bm P}_{*m}$, and we assume that three-particle intermediate states with on-shell kinematics are not possible given the initial-state energy. The problematic contributions satisfying conditions $i$ and $ii$ can be subtracted from Eq.~(\ref{eq:GLE}), leaving the rest to read
\begin{align}
\mathcal{T}_L^{(\rm{E})\,\geq} \equiv \int d\tau \,e^{E_1\tau} \left[ G_L^{(\rm E)}(\tau)-G_L^{(\rm{E})<}(\tau) \right].
\label{eq:TLElarger}
\end{align}
The spectral decomposition of $\mathcal{T}_L^{(\rm{E})\,\geq}$ has, therefore, exactly the same form as the Minkowski counterpart upon an overall $i$ factor. Here,
\begin{align}
G_L^{(\rm{E})<}(\tau) \equiv \sum_{m=0}^{N-1} \theta(\tau) \, c_m \, e^{-(|\bm{P}_{*m}|+E_{*m}-E_f)|\tau|}+
\nonumber\\
\sum_{m=0}^{N'-1} \theta(-\tau) \, c_{m} \, e^{-(|\bm{P}_{*m}|+E_{*m}-E_i)|\tau|},
\label{eq:}
\end{align}
where it is assumed that there are $N~(N')$ states satisfying condition $i~(ii)$ above, and
\begin{equation}
c_m \equiv \frac{m_{\beta\beta}}{2|\bm{P}_{*m}|}  \big [\langle E_f,L|\mathcal{J}(0)| E_{*m},L \rangle \langle E_{*m},L|\mathcal{J}(0) |E_i,L \rangle \big]_L.
\label{eq:cm}
\end{equation}

The remaining contributions arising from on-shell intermediate states, called $\mathcal{T}_L^{(\rm{E})\,<}$, can be formed separately with the knowledge of the single-current matrix elements in a finite volume between the initial (final) and intermediate states:
\begin{eqnarray}
\mathcal{T}_L^{(\rm{E})\,<} \equiv \sum_{m=0}^{N-1} \frac{c_m}{|\bm{P}_{*m}|+E_{*m}-E_f-E_1}+
\nonumber\\
\sum_{m=0}^{N'-1} \frac{c_{m}}{|\bm{P}_{*m}|+E_{*m}-E_i+E_1}.
\label{eq:TLEsmaller}
\end{eqnarray}

Equations~(\ref{eq:TLEsmaller}) and (\ref{eq:TLElarger}) can now be combined to construct the desired Minkowski quantity $\mathcal{T}_L^{(\rm M)}$,
\begin{equation}
\mathcal{T}_L^{(\rm M)} =i\mathcal{T}_L^{(\rm{E})\,<}+i\mathcal{T}_L^{(\rm{E})\,{\geq}},
\label{eq:TLM}
\end{equation}
whose relation to the physical $nn \to pp \;(ee)$ amplitude at LO in the EFT was already established in Eq.~(\ref{eq:IVFVmatching}). This completes the matching  framework that relates $\mathcal{M}^{(\rm{Int.})}_{nn\to pp}$ in Eq.~(\ref{eq:MICnnpp}), and hence the new short-distance LEC $g_\nu^{NN}$, to the LQCD correlation function $G_L^{(\rm E)}(\tau)$ in Eq.~(\ref{eq:GLE}). It should be noted that the single-current matrix elements required for this matching relation, i.e., those appearing in Eq.~(\ref{eq:cm}), can themselves be evaluated with LQCD, and can be matched to the physical amplitude for the single-$\beta$ transition amplitude~\cite{Briceno:2012yi, Davoudi:2020xdv}.

\vspace*{3mm}
{\it Discussion and outlook.---}Given significant progress in LQCD studies of nuclear matrix elements in recent years~\cite{Beane:2015yha, Savage:2016kon, Shanahan:2017bgi, Winter:2017bfs, Chang:2017eiq, Drischler:2019xuo, Detmold:2020snb, Davoudi:2020ngi}, albeit yet with unphysical quark masses, it is expected that LQCD will be able to evaluate the four-point correlation function in Eq.~(\ref{eq:GLE}), along with the required two- and three-point functions that allow the construction of the finite-volume Minkowski amplitude in Eq.~(\ref{eq:TLM}). This can then be used in Eq.~(\ref{eq:IVFVmatching}) to constrain the physical EFT amplitude, hence the unknown short-distance contribution. The practicality of the method, however, relies on the presence of only a finite (and few) number of on-shell intermediate states that are composed of no more than two hadrons. One can estimate the expected nature and the number of intermediate states by examining a plausible example. Let us take $L=8$ fm to ensure the validity of the finite-volume formalism used with physical quark masses, up to exponentially suppressed contributions~\cite{Sato:2007ms, Briceno:2013bda}. The finite-volume spectrum of the two-nucleon isotriplet channel at rest arising from singularities of the function in Eq.~(\ref{eq:calF0}) can be determined using the experimentally known phase shifts~\cite{Stoks:1994wp}, giving the ground-state energy $E_{n_i}  \approx -2.6$ MeV (which polynomially approaches zero as $L \to \infty$). A simple kinematic can be considered for the transition amplitude such that $E_{i}(=E_{n_i})=E_f(=E_{n_f})$, and where the currents carry zero energy and momentum so that the final-state two-nucleon system remains at rest. Given the available total energy, and the quantum numbers of the currents, the only allowed intermediate state is the two-nucleon isotriplet channel at rest, whose low-lying spectrum in this volume is $\widetilde{E}_{*m} \approx \{-5.6,13.9,\cdots\}$ MeV. While it may appear that the ground state of this system constitutes an on-shell intermediate state, requiring construction of the Minkowski amplitude through an evaluation of the isosinglet to isotriplet matrix element, one must note that since the zero spatial momentum is not allowed for the neutrino propagation in the finite volume, none of the on-shell conditions stated before can be satisfied with the kinematics considered (noting that the minimum allowed energy of an on-shell neutrino in this volume is $|\bm{P_*}|=2\pi/L \approx 155.0$ MeV). As a result, $\mathcal{T}_L^{(\rm M)} =i\mathcal{T}_L^{(\rm{E})}=i\int d\tau \,e^{E_1\tau} G_L^{(\rm E)}(\tau)$, and Eq.~(\ref{eq:IVFVmatching}) can be readily used to obtain the physical amplitude from the LQCD four-point function $G_L^{(\rm E)}(\tau)$. 

This example demonstrates that obtaining the physical amplitude of the $nn \to pp\;(ee)$ process from LQCD is even more straightforward than its two-neutrino counterpart, as in the latter there is a larger kinematic phase space allowed for on-shell intermediate states. The current framework, therefore, takes an essential step in enabling constraints on $g_\nu^{NN}$ directly from LQCD in the upcoming years. Besides its application in the $0\nu\beta\beta$ process, the formalism outlined will find its use in a range of hadronic processes that consist of single- or two-hadron initial, intermediate, and final states, and where a light lepton (or photon) propagator is present, such as in the semi-leptonic rare decays of the kaon~\cite{Christ:2020hwe}, and the virtual-photon contribution to charge-invariance breaking in the two-nucleon sector~\cite{Epelbaum:1999zn,Cirigliano:2020dmx}.

\vspace*{3mm}
\noindent
{\it Acknowledgments.---}ZD acknowledges valuable discussions with William Detmold during the initial stage of this work at Massachusetts Institute of Technology. ZD would like to further thank members of the NPLQCD Collaboration, in particular Martin Savage, for numerous fruitful discussions regarding the topic of double-$\beta$ decays from LQCD. She further appreciates insightful discussions and collaborations with Ra\'ul Brice\~no, Maxwell Hansen, and Matthias Schindler on the topic of long-range hadronic matrix elements. ZD and SVK are supported by the Alfred P. Sloan fellowship, and by the Maryland Center for Fundamental Physics at the University of Maryland, College Park. ZD is further supported by the U.S. Department of Energy's Office of Science Early Career Award, under award no. DE-SC0020271. The work performed at Massachusetts Institute of Technology was supported by the U.S. Department of Energy Early Career Award DE-SC0010495 and grant number DE-SC0011090

\bibliography{bibi.bib}

\appendix
\begin{widetext}
\section{Supplemental Material
\label{app:supp}}
\noindent
The steps involved in performing the two-loop sum-integral difference defined in Eq.~(9) of the main text for an expedited convergence will be outlined in this section. To simplify the notation, the conventions $n \equiv |\bm{n}|$ and $n^2 \equiv |\bm{n}|^2$ are used for any three-vector $\bm{n}$.

The two-loop integral involving the neutrino propagator is given by
\begin{equation}
\label{eq: definition J(epsilon) in inf V}
J^{\infty}(p_1^2,p_2^2) =
M^2 \int \frac{d^3k_1}{(2\pi)^3} \, \frac{d^3k_2}{(2\pi)^3}
\frac{1}{p_1^2-k_1^2+i\epsilon} \, \frac{1}{p_2^2-k_2^2+i\epsilon} \, \frac{1}{|\bm{k}_1 - \bm{k}_2|^2}.
\end{equation}
This integral is divergent in the UV region of integrating variables and must be regulated. While for the discussion of the physical amplitude in the main text, dimensional regularization is a natural choice as presented in Eq.~(6) of the main text~\cite{Cirigliano:2019vdj}, for the evaluation of the sum-integral difference, a cutoff regulator  $\Lambda$ proves most useful. As the physical amplitude, as well as the matching condition are UV convergent, both choices can be used in the matching framework. In particular, with the cutoff regularization, $J^{\infty}$ evaluates to
%
\begin{eqnarray}
J^\infty(p_1^2,p_2^2)  &=&
\frac{M^2}{8\pi^4}
\int_{0}^{\Lambda} dk_1 \, 
\int_{0}^{\infty}  dk_2 \,
\frac{k_1^2}{p_1^2-k_1^2} \, \frac{k_2^2}{p_2^2-k_2^2}
\int_{-1}^{1} dx \, \frac{1}{k_1^2+k_2^2-2\,k_1\,k_2\,x}
+ \frac{iM^2}{32\pi} - \frac{M^2}{32 \pi^2} \ln\Bigg(\frac{p_1+p_2}{|p_1-p_2|}\Bigg)\nonumber\\
&=&M^2 \bigg[
\frac{\ln{\Lambda}}{16\pi^2} + \frac{i}{32\pi} - \frac{\ln{(p_1+p_2)}}{16 \pi^2}\bigg].
\label{eq: cutoff regulated inf V}
\end{eqnarray}

In a finite volume with cubic geometry and spatial extent $L$ along each Cartesian coordinate and with periodic boundary conditions, the analog of Eq.~\eqref{eq: definition J(epsilon) in inf V} is given by replacing integrals with sums over quantized three-momenta $\bm{k}=2\pi\bm{n}/L$ with $\bm{n} \in \mathbb{Z}^3$:
\begin{align}
\label{eq: definition J in fin V}
J^V(p_1^2,p_2^2)&=\frac{M^2}{L^6}
\sum_{\substack{
\bm{k}_1\\ 
k_1^2\neq p_1^2}}
\sum_{\substack{
\bm{k}_2\neq\bm{k}_1\\ 
k_2^2\neq p_2^2}}
\frac{1}{p_1^2-k_1^2} \, \frac{1}{p_2^2-k_2^2} \, \frac{1}{|\bm{k}_1 - \bm{k}_2|^2} .
\end{align}
Here, the $i\epsilon$ terms are dropped from the denominators since discrete sums are defined over non-singular values of $\bm{k}_1$ and $\bm{k}_2$.

Equation~\eqref{eq: definition J in fin V} differs from Eq.~\eqref{eq: definition J(epsilon) in inf V} by power-law correction in $1/L$ which can be isolated from the difference
\begin{equation}
\label{eq: S FV as S(epsilon) and delta S}
 \delta J^V(p_1^2,p_2^2) \equiv J^V(p_1^2,p_2^2) - J^\infty(p_1^2,p_2^2) \; .
\end{equation}
To evaluate $\delta J^V$, let us first convert the summation variable in Eq.~\eqref{eq: definition J in fin V} from $\bm{k}_{1(2)}$ to $\bm{n}_{1(2)}$ and rescale $p_{1(2)}$ as $\tilde{p}_{1(2)}=p_{1(2)}\,L/2\pi$. Next, one can observe that the UV divergence in Eq.~\eqref{eq: definition J in fin V} is the same as that occurred in the sum when $p_1^2=p_2^2=0$. Using a cutoff regulator $\Lambda$, this sum reads
\begin{equation}
\label{eq: cutoff regulated in FV}
J^V(0,0)=\frac{M^2}{(2\pi)^6}
\sum_{\bm{n}_1\neq 0}^{\tilde{\Lambda}} \;
\sum_{\bm{n}_2\neq 0,\bm{n}_1}
\frac{1}{n_1^2} \, \frac{1}{n_2^2} \, \frac{1}{|\bm{n}_1 - \bm{n}_2|^2}.
\end{equation}
The upper bound on the sum over $\bm{n}_1$ indicates that only integer triplets that satisfy $n_1 \leq \tilde{\Lambda}\,(=\Lambda L/2\pi)$ must be included. The sum over $\bm{n}_2$ is left unbounded. Now adding and subtracting $J^V(0,0)$ and upon using Eq.~\eqref{eq: cutoff regulated inf V}, Eq.~\eqref{eq: S FV as S(epsilon) and delta S} becomes
\begin{equation}
\label{eq: definition delta S in R X3 X6}
\delta J^V(p_1^2,p_2^2) = \frac{M^2}{(2\pi)^6}\bigg[ 
\mathcal{R}
-\mathcal{X}_1(\tilde{p}_1^2,\tilde{p}_2^2)
-\frac{\mathcal{X}_3(\tilde{p}_2^2)}{\tilde{p}_1^2}
-\frac{\mathcal{X}_3(\tilde{p}_1^2)}{\tilde{p}_2^2}
 + \mathcal{X}_6(\tilde{p}_1^2,\tilde{p}_2^2)
\bigg]
+ \frac{M^2}{16\pi^2}\ln{(\tilde{p}_1+\tilde{p}_2)}
-\frac{iM^2}{32\pi} \; ,
\end{equation}
where
\begin{align}
\label{eq: definition R}
&\mathcal{R} \equiv \lim\limits_{\tilde{\Lambda}\to\infty} \Bigg[ \;
\sum_{\bm{n}_1\neq 0}^{\tilde{\Lambda}} \;
\sum_{\bm{n}_2\neq 0,\bm{n}_1}
\frac{1}{n_1^2} \, \frac{1}{n_2^2} \, \frac{1}{|\bm{n}_1 - \bm{n}_2|^2}
- 4\pi^4\ln{\tilde{\Lambda}}
\Bigg] ,
\\
& \mathcal{X}_1(\tilde{p}_1^2,\tilde{p}_2^2)=
\frac{1}{\tilde{p}_1^2\,\tilde{p}_2^2}
\sum_{\substack{
\bm{n}_1\\
n_1^2=\tilde{p}_1^2
}
} \;
\sum_{\substack{
\bm{n}_2\neq \bm{n}_1\\
n_2^2=\tilde{p}_2^2
}
}\frac{1}{|\bm{n}_1- \bm{n}_2|^2},
\\
\label{eq: definition X3}
&\mathcal{X}_3 (\tilde{p}) \equiv
\sum_{\substack{
\bm{n}\neq 0\\ 
n^2\neq \tilde{p}^2}} \,
\frac{1}{n^2(n^2-\tilde{p}^2)},
\\
\label{eq: definition X6}
& \mathcal{X}_6 (\tilde{p}_1^2,\tilde{p}_2^2) \equiv 
\sum_{\substack{
\bm{n}_1\neq 0\\ 
n_1^2\neq \tilde{p}_1^2}}
\sum_{\substack{
\bm{n}_2\neq 0,\bm{n}_1\\ 
n_2^2\neq \tilde{p}_2^2}}
\bigg[\frac{1}{\tilde{p}_1^2-n_1^2} \, \frac{1}{\tilde{p}_2^2-n_2^2} 
-\frac{1}{n_1^2} \, \frac{1}{n_2^2}\bigg]\, \frac{1}{|\bm{n}_1
- \bm{n}_2|^2}.
\end{align}

To evaluate the lattice sums in Eqs.~\eqref{eq: definition R}-\eqref{eq: definition X6}, one can use the method of tail-singularity separation (TSS) described in Ref.~\cite{Tan:2007bg}. In this method, the sum is split into two pieces: one containing the singular contributions and the other containing a power-law tail which is sufficiently smooth such that it can be approximated by its integral counterpart. As an example, let us sketch out the details for evaluating the lattice sum in Eq.~\eqref{eq: definition X3}. The TSS scheme can be achieved by introducing exponential factors containing a small positive number, $\alpha$, and rewriting $\mathcal{X}_3$ as
\begin{equation}
\mathcal{X}_3(\tilde{p}^2) \, = \, 
\sum_{\substack{
\bm{n}\neq 0\\ 
n^2\neq \tilde{p}^2}} \,
\frac{\big[e^{-\alpha n^2}+1-e^{- \alpha n^2}\big]}{n^2} \frac{\big[e^{-\alpha(n^2-\tilde{p}^2)}+1-e^{-\alpha(n^2-\tilde{p}^2)}\big]}{n^2-\tilde{p}^2} .
\end{equation}
The smooth function containing a power-law tail is obtain by gathering $\big[1-e^{-\alpha n^2}\big]$ and $\big[1-e^{-\alpha (n^2-\tilde{p}^2)}\big]$ factors, which is then approximated, up to $\mathcal{O}(e^{-\pi^2/\alpha})$ corrections, by $\sum_{\bm{n}}\to\int d^3n$ with values at the poles removed. This gives
\begin{align}
\nonumber
\mathcal{X}_3(\tilde{p}^2) \, &= \, 
\sum_{\substack{
\bm{n}\neq 0\\ 
n^2\neq \tilde{p}^2}}
\frac{\big[e^{-\alpha(n^2-\tilde{p}^2)}+ e^{-\alpha n^2}-e^{-\alpha(2n^2-\tilde{p}^2)}\big]}{n^2(n^2-\tilde{p}^2)}\\
&+ \int d^3n \; \frac{\big[1-e^{-\alpha n^2}\big]\big[1-e^{-\alpha(n^2-\tilde{p}^2)}\big]}{n^2(n^2-\tilde{p}^2)}
-\frac{2\alpha}{\tilde{p}^2}\sinh{(\alpha \tilde{p}^2)} + \mathcal{O}(e^{-\pi^2/\alpha}).
\label{eq: TSS on X3}
\end{align}
The convergence is obtained as $\alpha\to0^+$ and the converged value is independent of $\alpha$ up to exponential corrections in $1/\alpha$. For $\tilde{p}=1$ and $\alpha$ as large as 0.1, Eq.~\eqref{eq: TSS on X3} converges to $\mathcal{X}_3(1)=14.7$. $\alpha=0.01$ gives $\mathcal{X}_3(1)=14.702$, which is in agreement with Eq.~(A1) of Ref.~\cite{Beane:2014qha} up to five significant figures.

This method can be extended to double sums as demonstrated in Ref.~\cite{Tan:2007bg}. $\mathcal{R}$ defined in Eq.~\eqref{eq: definition R} has been evaluated using TSS in Eq.~(30) of Ref.~\cite{Beane:2014qha}:
\begin{equation}\label{eq: value of R}
\mathcal{R}=-178.42 .
\end{equation}

For the case of $\tilde{p}_1=\tilde{p}_2=1$, $\mathcal{X}_1$ in Eq.~\eqref{eq: definition X6} straightforwardly evaluates to $27/2$. On the other hand, $\mathcal{X}_6$ can be rewritten as
\begin{equation}
\label{eq: X6 expanded}
\mathcal{X}_6(\tilde{p}^2,\tilde{p}^2) =  
\, \sum_{\substack{
\bm{n}_1\neq 0\\ 
n_1^2\neq \tilde{p}^2}}
\,\frac{2 \, \tilde{p}^2\,\mathcal{X}_6^{(1)}(\bm{n}_1,\tilde{p}^2)-\tilde{p}^4\,\mathcal{X}_6^{(2)}(\bm{n}_1,\tilde{p}^2)}
{n_1^2\,(n_1^2-\tilde{p}^2)},
\end{equation}
with
\begin{align}
\label{eq: definition X6-1 and X6-2}
\mathcal{X}_6^{(1)}(\bm{n}_1,\tilde{p}^2) = 
\sum_{\substack{
\bm{n}_2\neq 0,\bm{n}_1\\ 
n_2^2\neq \tilde{p}^2}}
\frac{1}{(n_2^2-\tilde{p}^2)|\bm{n}_1-\bm{n}_2|^2} ,
\qquad
\mathcal{X}_6^{(2)}(\bm{n}_1,\tilde{p}^2) =
\sum_{\substack{
\bm{n}_2\neq 0,\bm{n}_1\\ 
n_2^2\neq \tilde{p}^2}}
\frac{1}{n_2^2\,(n_2^2-\tilde{p}^2)|\bm{n}_1-\bm{n}_2|^2}.
\end{align}
TSS can now be used on $\mathcal{X}_6^{(1)}(\bm{n}_1,\tilde{p}^2)$ and $\mathcal{X}_6^{(2)}(\bm{n}_1,\tilde{p}^2)$ in Eq.~\eqref{eq: definition X6-1 and X6-2} just like in Eq.~\eqref{eq: TSS on X3}:
\begin{align}
\nonumber
\mathcal{X}_6^{(1)}(\bm{n}_1,\tilde{p}^2) &=
\sum_{\substack{
\bm{n}_2\neq 0,\bm{n}_1\\ 
n_2^2\neq p^2}}
\frac{1}{(n_2^2-\tilde{p}^2)|\bm{n}_1-\bm{n}_2|^2}\big[e^{-\alpha |\bm{n}_1-\bm{n}_2|^2}+e^{-\alpha (n_2^2-\tilde{p}^2)}-e^{-\alpha (|\bm{n}_1-\bm{n}_2|^2+n_2^2-\tilde{p}^2)}\big]
\\
\label{eq: TSS for X6-1}
&+\mathcal{P}\,\int\,d^3n_2 \, 
\frac{\big[1-e^{-\alpha |\bm{n}_1-\bm{n}_2|^2}\big]}{|\bm{n}_1-\bm{n}_2|^2}
\frac{\big[1-e^{-\alpha (n_2^2-\tilde{p}^2)}\big]}{n_2^2-\tilde{p}^2}
-\alpha\frac{\big[1-e^{-\alpha (n_1^2-\tilde{p}^2)}\big]}{n_1^2-\tilde{p}^2}
+\mathcal{O}(e^{-\pi^2/\alpha}),\\
\nonumber
\mathcal{X}_6^{(2)}(\bm{n}_1,\tilde{p}^2) &=
\sum_{\substack{
\bm{n}_2\neq 0,\bm{n}_1\\ 
n_2^2\neq \tilde{p}^2}}
\frac{1}{n_2^2(n_2^2-\tilde{p}^2)|\bm{n}_1-\bm{n}_2|^2}
\bigg[ e^{-\alpha |\bm{n}_1-\bm{n}_2|^2}
+ e^{-\alpha (n_2^2-\tilde{p}^2)}+e^{-\alpha n_2^2}
\\
\nonumber
& + e^{-\alpha (|\bm{n}_1-\bm{n}_2|^2+2n_2^2-\tilde{p}^2)}-e^{-\alpha( |\bm{n}_1-\bm{n}_2|^2+n_2^2)}
-e^{-\alpha (|\bm{n}_1-\bm{n}_2|^2+n_2^2-\tilde{p}^2)}
-e^{-\alpha (2n_2^2-\tilde{p}^2)}
\bigg]
\\
\nonumber
&+\mathcal{P}\,\int\,d^3n_2 \, 
\frac{\big[1-e^{-\alpha |\bm{n}_1-\bm{n}_2|^2}\big]}{|\bm{n}_1-\bm{n}_2|^2}
\frac{\big[1-e^{-\alpha n_2^2}\big]}{n_2^2}
\frac{\big[1-e^{-\alpha (n_2^2-\tilde{p}^2)}\big]}{n_2^2-\tilde{p}^2}\\
\label{eq: TSS for X6-2}
&+\alpha\frac{\big[1-e^{\alpha \tilde{p}^2}\big]}{\tilde{p}^2}\frac{\big[1-e^{-\alpha n_1^2}\big]}{n_1^2}
-\alpha\frac{\big[1-e^{-\alpha (n_1^2-\tilde{p}^2)}\big]}{n_1^2-\tilde{p}^2}\frac{\big[1-e^{-\alpha n_1^2}\big]}{n_1^2}
+\mathcal{O}(e^{-\pi^2/\alpha}).
\end{align}
Here, $\mathcal{P}$ denotes the Cauchy principal value of the radial integration in $\bm{n}_2$ for the pole $n_2^2=\tilde{p}^2$. Using Eqs.~\eqref{eq: TSS for X6-1} and \eqref{eq: TSS for X6-2} in Eq.~\eqref{eq: X6 expanded}, the outer sum over $\bm{n}_1$ is then split into two parts: terms containing exponentially suppressed terms in $\alpha$ and terms independent of $\alpha$. The former can be summed directly while the latter, which stems from evaluating the $\alpha$-independent integrals in Eq.~\eqref{eq: TSS for X6-1} and \eqref{eq: TSS for X6-2}, can be calculated using TSS just like Eq.~\eqref{eq: TSS on X3}. As an example, with this procedure, Eq.~\eqref{eq: X6 expanded} for $\tilde{p}^2=1$ evaluates to
\begin{equation}
\mathcal{X}_6(1,1) = 264,
\end{equation}
which agrees with Eq.~(A10) of Ref.~\cite{Beane:2014qha} up to three significant figures. Arbitrary accuracy can be achieved by decreasing the value of $\alpha$ and increasing the number of integer-triplets used in the sums with increasing magnitude.

\end{widetext}

\end{document}